\documentclass{ifacconf}

\usepackage{graphicx}      
\usepackage{natbib}        
\usepackage{bm}
\usepackage{amsmath}
\DeclareMathOperator*{\argmax}{arg\,max}
\DeclareMathOperator*{\argmin}{arg\,min}

\begin{document}
\begin{frontmatter}

\title{Fitting nonlinear models to continuous oxygen data with oscillatory signal variations via a loss based on Dynamic Time Warping \thanksref{footnoteinfo}} 

\thanks[footnoteinfo]{This work was supported by the German Federal Ministry of Education and Research through the Program International Future Labs for Artificial Intelligence (Grant number 01DD20002A).}

\author[First]{Judit Aizpuru}
\author[First]{Annina Karolin Kemmer}
\author[First]{Jong Woo Kim}
\author[First]{Stefan Born}
\author[First]{Peter Neubauer}
\author[First,Second]{Mariano N. Cruz Bournazou}
\author[Third]{Tilman Barz} 

\address[First]{Technische Universit{\"a}t Berlin, Chair of Bioprocess Engineering, Strasse des 17. Juni 135, 10623 Berlin, Germany}
\address[Second]{DataHow AG, Z{\"u}richstrasse 137, 8600 D{\"u}bendorf, Switzerland}
\address[Third]{AIT Austrian Institute of Technology GmbH, Center for Energy, Giefingasse 2, 1210 Vienna, Austria}

\begin{abstract}                
    High throughput experimental systems play an important role in bioprocess development, as they provide an efficient way of analysing different experimental conditions and perform strain discrimination in previous phases to the industrial scale production. In the millilitre scale, these systems are combinations of parallel mini-bioreactors, liquid handling robots and automated workflows for data handling and model based operation. For successfully monitoring cultivation conditions and improving the overall process quality by model-based approaches, a proper model identification is crucial. However, the quality and amount of measurements makes this task challenging considering the complexity of the bio-processes. The Dissolved Oxygen Tension is often the only measurement which is available online, and therefore, a good understanding of the errors in this signal is important for performing a robust estimation. Some of the expected errors will provoke uncertainties in the time-domain of the measurement, and in those cases, the common Weighted Least Squares estimation procedure can fail providing good results. Moreover, these errors will have even a larger effect in the fed-batch phase where bolus feeding is applied, as this generates fast dynamic responses in the signal. In the present work, an insilico study of the performance of Weighted Least Squares estimator is analysed when the expected time-uncertainties are present in the oxygen signal. As an alternative, a loss based on the Dynamic Time Warping measure is proposed. The results show how this latter procedure outperforms the former reconstructing the oxygen signal, and in addition, returns less biased parameter estimates.
    
\end{abstract}


\end{frontmatter}

\section{Introduction}

Characterising process models in biotechnology is still a very challenging task.
Apart from the general difficulties in parameter 
estimation for nonlinear models and model-specific problems
with correlations between parameters, the reduced availability and 
quality of the measurements is the major obstacle to overcome.
The measurements of the principal states contributing in the identification of the model are generally scarce and require several processing and calibration steps to finally be usable. Yet, the Dissolved Oxygen Tension (DOT) is often monitored online, as it contains important information about the cellular activity in the bioreactor \citep{ruffieux_measurement_1998} and usually is a limiting factor in the cultivations \citep{suresh_techniques_2009}. Several continuous measurement techniques have already been developed \citep{wang_optical_2014, flitsch_easy_2016}, and among these, optical sensors based on fluorescent quenching are the most suited for small scale bioreactors, as they provide a non-invasive technique which does not consume oxygen and allows easy miniaturisation \citep{wei_review_2019}.

The identification of the model will highly depend on the fit of the oxygen signal and it is important to understand the sources of errors that can happen in order to perform a correct estimation. One of the limitations of its online measurement is that the response time of the optical sensors is not immediate. The reported sensor response times range from ${10\! -\!40 }$ s \citep{achatz_luminescent_2011}\footnote{https://www.presens.de} for sensor spots consisting in a fluorescent dye. Approaches to analyse this problem and proposals for new sensing techniques with lower response times can be found in the literature \citep{flitsch_easy_2016}, however, using sensor spots is still a more usual practice. Other possible sources of errors may result from wrong assumptions in the modelling of the oxygen mass transfer, i.e. when using models with constant volumetric oxygen transfer rates (kLa) during
the entire fermentation. For example, neglecting changing biomass concentrations and media compositions can provoke unmodelled expansions
or contractions of the signal \citep{pappenreiter_oxygen_2019}. In addition, in cultivations where the feeding is performed in bolus-mode, the inputs to the system might experience some delays due to the viscosity of the feed, its volume or the mixing times in the dissolution, producing a mismatch between the deterministic inputs to the simulation and the actual times when these occur. What all these errors have in common is, that they produce uncertainties in the time-domain of the measurement, and when abrupt changes of oxygen happen, these might produce large residuals which violate the assumptions of the common estimation procedure. 

The most common procedure for the identification of parametric models is the Weighted Least Squares (WLS) estimation procedure. This can be seen as a maximum likelihood estimation with independent normally distributed measurement errors in the dependent variables, where errors in the independent variables, usually \textit{time} for dynamic models, are not taken into account. Furthermore, from the studies on robust statistics \citep{huber_robust, ozyurt_theory_2004, de_menezes_review_2021, da_cunha_robust_2021}, it is well known that the least squares estimator is not robust against large errors violating the normality and independence assumptions. It can be expected that the existing time uncertainties will provoke both scenarios. 

An alternative measure of similarity between time series that seems to be appropriate for coping with the mentioned uncertainties, is the Dynamic Time Warping (DTW). This measure is well known in the machine learning community due to its robustness for quantitatively comparing different length time series which can be shifted, dilated or contracted in time. Since its first application for speech recognition \citep{sakoe_hiroaki_dynamic_1978}, several variants and applications can be found in the literature, mostly related to pattern recognition and time series classification and clustering \citep{keogh_derivative_nodate, eilers_parametric_2004, cuturi_soft-dtw_2018, blondel_dierentiable_nodate, clustering-review, jeong_weighted_2011}.

The specific case study that is addressed in this work, is the cultivation of \textit{Escherichia Coli} in an automatic high throughput (HT) parallel cultivation system where bolus feeding is applied. The experimental metadata, measurements, set-points, and control inputs are measured and updated in a central database, which allows automatic operation as well as online model-based bioprocess monitoring and control \citep{cruz_bournazou_online_2017}. The DOT is measured using fluorescent dye sensor spots integrated at the bottom of the mini-bioreactors and with a sampling period of approximately 60s. The used bolus feeding strategy produces fast oxygen responses to the sudden substrate concentration changes, which produces a pattern-like oscillatory signal of oxygen. Thus, small time uncertainties in the measurements can produce large residuals between the data and the simulation, and therefore the most informative measurement for characterising the process model can spoil the good behaviour of the common estimation method whenever these time uncertainties are present. 

In the present work, an analysis of the performance of the common WLS estimator is presented considering DOT data where the expected time uncertainties are generated artificially. As an alternative fitting methodology, a differentiable divergence between time series based on the DTW is proposed as a loss function. For doing so, a synthetic dataset is generated in silico using the macrokinetic growth model for \textit{E. coli}, presented in \citep{anane_modelling_2019}. The data is generated containing delayed inputs of bolus feed, which reflects the difficult to model time shifts due to sensor delay, non-ideal mixing, and actual uncertainties in the feeding times.

\section{Methods}

This section present the two ways used for quantifying the goodness of the fit of a given model to the experimental data, formulating the parameter estimation procedure as an optimisation problem. 


\subsection{Weighted Least Squares}
\label{sec:WLS}
Considering our system model to be a function:

\begin{equation}
   \boldsymbol{y} = f(t, \boldsymbol{u}, \boldsymbol{ \theta } )
\end{equation}

The errors or residuals ($\boldsymbol{\epsilon}$) between measurements ($\boldsymbol{y}^m$) and model predictions ($\boldsymbol{y}$) are described as:

\begin{equation}
   \epsilon_{i,j} =  y^m_{i,j} - y_{i,j} 
\end{equation}

Where $i=1,..,n_s$ represents the states and $j=1,...,n_t^i$ the time stamps of the measurements.

In the optimal parameter values, these residuals are assumed to follow a normal distribution centred at zero and with standard deviation equal to the measurement accuracy $\sigma_i$. Thus, the probability of each individual residual under the assumed distribution, will be given by the probability density function in equation \ref{eq:pdf}.

\begin{equation}
\label{eq:pdf}
    f(\epsilon_{i,j}|0,\sigma_i) = \frac{1}{ \sigma_i \sqrt{2\pi}} 
    exp \left(- \frac{1}{2} \frac{\epsilon_{i,j}^2}{\sigma_i^2} \right)
\end{equation}

Considering the errors are independent from each other, the inference of the model parameters can be seen as a maximisation problem, where the optimal parameters will maximise the probability of getting the assumed residual distribution:

\begin{equation}
    \boldsymbol{\theta^*} = \argmax_{\boldsymbol{\theta}} \prod_{i=1}^{n_s} \prod_{j=1}^{n_t} f(\epsilon_{i,j}(\boldsymbol{\theta}))
\end{equation}

By changing the sign of the expression and taking its logarithm, the simplified expression is the commonly used WLS estimation objective:

\begin{equation}
    \label{OF_WLS}
    \Phi_{WLS} = \min_{\boldsymbol{\theta}}  \sum_i \sum_j \frac{1}{2} \frac{\epsilon_{i,j}^2}{\sigma_{i}^2}
\end{equation}

Where $\sigma_{i}$ is the accuracy of the measurement of state $i$ and the minimised function the WLS objective. This estimation procedure will return an asymptotically unbiased estimate of the model parameters when the assumptions are correct, which means that using a large amount of data, the estimates MLE will converge to the real parameter values. 


\subsection{Dynamic time warping}
\label{sec:DTW}

DTW is defined as the minimum cost alignment between two time series, and is efficiently computed using Bellman's recursion. Therefore, the measure will not only provide a time invariant cost scalar, but also the alignment that makes that cost optimal, providing extra information that will be useful in most of the cases. 

Defining the elements of the cost matrix between two time series $X = x_1,...,x_n$ and $Y=y_1,...,y_m$ as:

\begin{equation}
    \boldsymbol{C}_{ij}(\boldsymbol{X},\boldsymbol{Y}) = || x_i - y_j ||_2^2
\end{equation}

Where the norm will be considered to be the squared L2 norm and the two time series will represent measurement data and simulation in this specific context. The DTW finds the warping path or alignment matrix that minimises the cost between them, and this optimisation problem is formulated as:

\begin{equation}
    DTW(C) = \min_{A \in \mathcal{A}(n,m)} \langle A,C \rangle
\end{equation}

Where the optimal alignment matrix is:
\begin{equation}
    A^* = \argmin_{A \in \mathcal{A}(n,m)} \langle A,C \rangle
\end{equation}

The elements of $A$ are binary ($A_{ij} \in [0,1]$) and it represents the optimal alignment between the two time series. This alignment is not necessarily unique. 

From the original definition in \citep{sakoe_hiroaki_dynamic_1978}, the alignment matrix or indifferently, the warping path, will fulfil:

\begin{itemize}
    \item \textbf{Monotonicity condition:} Forces the warping path to choose the indexes in an order preserving and increasing manner, as is natural for time series: 
    $$ i_{k-1} \leq i_k ~~ \text{and} ~~ j_{k-1} \leq j_k $$
    \item \textbf{Continuity condition:} Forces the warping path to have a match for all indexes
    $$ i_k - i_{k-1} \leq 1 ~~ \text{and} ~~ j_k - j_{k-1} \leq 1 $$
    \item \textbf{Boundary conditions:} The first and last indexes of the two time series will be matched:
    $$ w_1 = (1, 1) ~~ \text{and} ~~ w_k = (n, m) $$
\end{itemize}

\begin{figure}[h]
    \centering
    \includegraphics[width=\linewidth]{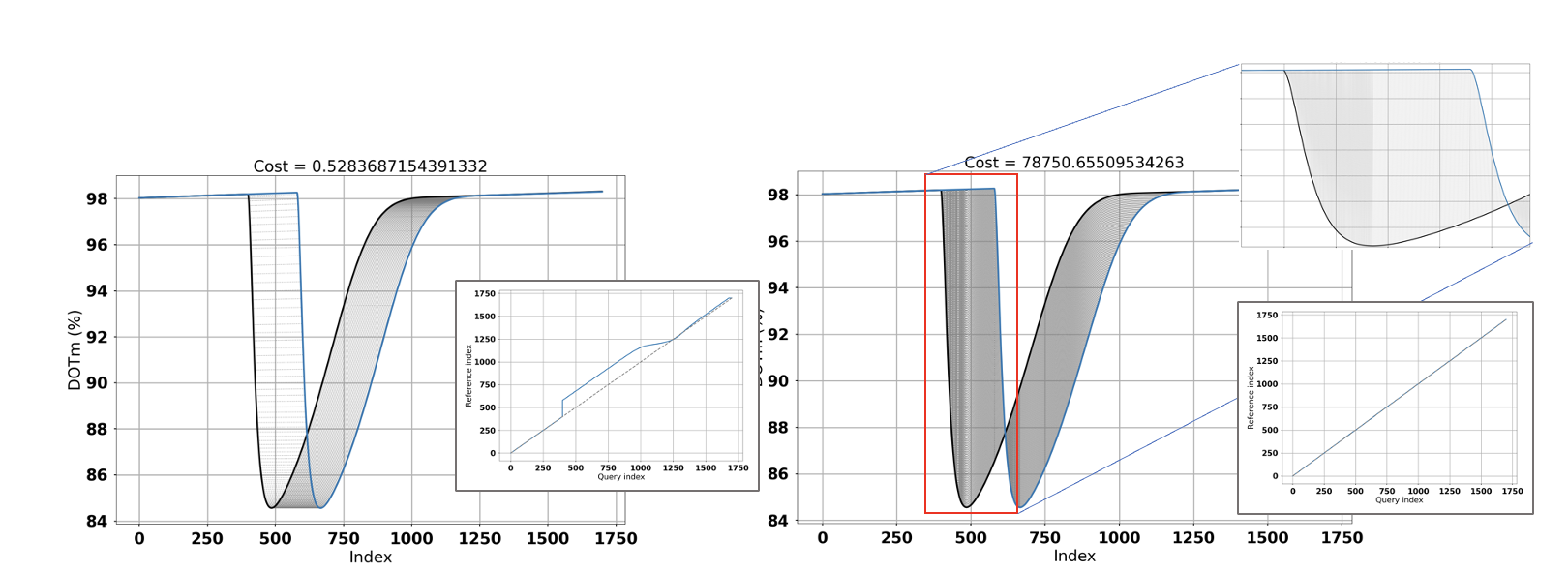} 
    \caption{DTW (left) and Euclidean (right) matching between two identical but time-shifted signals,
    and their corresponding cost and alignment matrixes}
    \label{fig:DTW_vs_Euc_matching}
\end{figure}

A variant of interest for our application is the differentiable version of this similarity measure. An approach to do so is presented in \citep{cuturi_soft-dtw_2018}, where the minimum operator is substituted by a soft minimum as:

\begin{equation}
    SDTW_{\gamma}(C) = \frac{1}{\gamma} \sum_{A \in \mathcal{A}(n,m)} exp \left( -\frac{\langle A,C \rangle}{\gamma} \right)
\end{equation}

Where $\gamma$ can be seen as the regularisation strength of the soft minimum operator. This new definition no longer returns an optimal alignment path, but an averaged version of it and the cost will be calculated among all those possibilities. 

Despite the success of the smoothed version of the DTW, one of its biggest drawbacks is that its value is not minimised when the two time series are the same. As a solution to this, the Soft-DTW divergence ($\mathrm{SDTWD}_{\gamma}$) is proposed in \citep{blondel_dierentiable_nodate}:

\begin{equation}
\begin{aligned}
    \label{OF_SDTWD}
    SDTWD_{\gamma}(X,Y) = SDTW_{\gamma}(X,Y) \\
    - \frac{1}{2}SDTW_{\gamma}(X,X) \\
    - \frac{1}{2}SDTW_{\gamma}(Y,Y)
\end{aligned}
\end{equation}

The authors also provide an open source Python package for the calculation of the SDTWD cost and its gradient, which can be found in Github\footnote{https://github.com/google-research/soft-dtw-divergences}.


\section{In silico experiment}
In order to analyse the performance of the two proposed objective functions under the expected errors, a synthetic dataset has been generated using the process model with fixed parameter values. The expected time uncertainties and the normal errors from the measurement systems precision are added in order to have a realistic scenario. 

\subsection{Model}
The model used for generating the data is described in \citep{anane_modelling_2019}. The nonlinear system of differential equations contains 7 state variables: X-biomass (g/L), S-substrate (g/L), A-acetate (g/L), DOTa-actual oxygen tension (\%), DOTm-measured oxygen tension (\%), V-volume (L) and P-product(g/L). The reaction kinetics and mass balance equations are described by 18 model parameters, which change depending on the strain and the experimental conditions, but certain bounds can be set using expert knowledge. In this case study, just 3 out of 18 parameters will be considered unknown, and their \textit{true} values and bounds are shown in Table \ref{tb:parameters}.

\begin{table}[h]
	\caption{True parameter values and bounds}
	\begin{center}
		\begin{tabular}{l c c} \hline
            Parameter      &   Value   &    Bounds   \\ \hline
            $q_{s,\mathrm{max}}$     &   1.60    &    1.20-1.70   \\
            $Y_{xs,\mathrm{em}}$     &   0.59    &    0.50-0.60    \\
            $k_{L}a$       &   373.6   &    200.0-700.0   \\
            \hline 
		\end{tabular} 
	\end{center}
	\label{tb:parameters}
\end{table}

\subsection{Synthetic data}
The generated synthetic data just contains oxygen data, as is the signal that will be more affected by the input-time uncertainties and is the only one that is available online. The sampling period for oxygen is $60$ s. Random normal errors with a standard deviation of $0.1$\%, as constant measurement accuracy, are added to the simulated oxygen signal. The experiment runs for a total of $8$ h, where a batch phase of approximately $3.5$ h will be followed by the fed-batch phase where the feed will be added in a bolus feeding-mode. The inputs to the system are the instantaneous and follow a discrete exponential feeding profile, with inputs every 10 min. The synthetic data is generated by perturbing the feed input time a constant amount of $60$ s, while the simulation to fit will contain the unperturbed feeding times. This way, the resulting oxygen response signal and the simulation for model identification are shifted imitating the possible sources of delay. The choice of using a constant delay of $60$s is to make the initial analysis simpler and use the maximum deviation which can be unperceived with the given sampling period.

\subsection{Numerical solution}

The differential equation system was solved using CVODES routine from the SUNDIALS suite \citep{hindmarsh_sundials_2005}, and the Assimulo package \citep{andersson_assimulo_2015} as an interface to it from the Python programming environment. In addition to the integrated states, the sensitivity solutions are also returned by the solver, which allows a more efficient computation of the gradients needed for solving the optimisation problem. Volume changes due to bolus feeds are modelled as instantaneous changes, and these are implemented using explicit time event functions.
With regard to the optimisation problem, 3 of the model parameters will be considered unknown. A random number within the parameter boundaries shown in \ref{tb:parameters} will be used as initial guess for the optimisation. As objective functions for minimisation, the WLS (eq. \ref{OF_WLS}) and SDTWD (eq. \ref{OF_SDTWD}) will be used, the latter with $\gamma = 0.1$. The optimisation algorithm is the \textit{L-BFGS-B} from Python's \textit{scipy} package, to which the gradients of the objective functions will be manually provided.

\subsection{Performance measures}
In order to compare the performance of the two objective functions when reconstructing the oxygen signal from the parameterised model, a different quantitative measure from the usual ones have to be defined. The usual measures, as the Root Mean Squared Error, cannot be used when one of the objective does not evaluate the solution as a point-wise difference where time is certain. Therefore, a new scoring function is defined based on the goodness of the solution in terms of biological meaning, where capturing the amplitude of the oxygen signal is considered to be of importance for proper control and correct oxygen uptake rate approximation.

The \textit{Amplitude Capturing Power} (ACP) is defined as a function of the Mean Absolute Error (MAE) of the moving maximum and minimum (Mmax, Mmin) between the experimental data and the simulation within a window \textit{w} as is shown in eq. \ref{eq:ACP}. The moving maxima (minima) of the signals is simply the maximum (minimum) values of the time series in a constant size rolling window along the entire time interval.

\begin{equation}
    \label{eq:ACP}
    ACP = f(MAE_{\mathrm{Mmax}_w}, MAE _{\mathrm{Mmin}_w})
\end{equation}

The MAE for the moving maximum is defined as:

\begin{equation}
    MAE_{\mathrm{Mmax}_w} = \frac{1}{n_w} \sum_{k=1}^{n_w} | Mmax_w(y^m)_k - Mmax_w(y)_k |
\end{equation}

\begin{figure}[h]
    \centering
    \includegraphics[width=\linewidth]{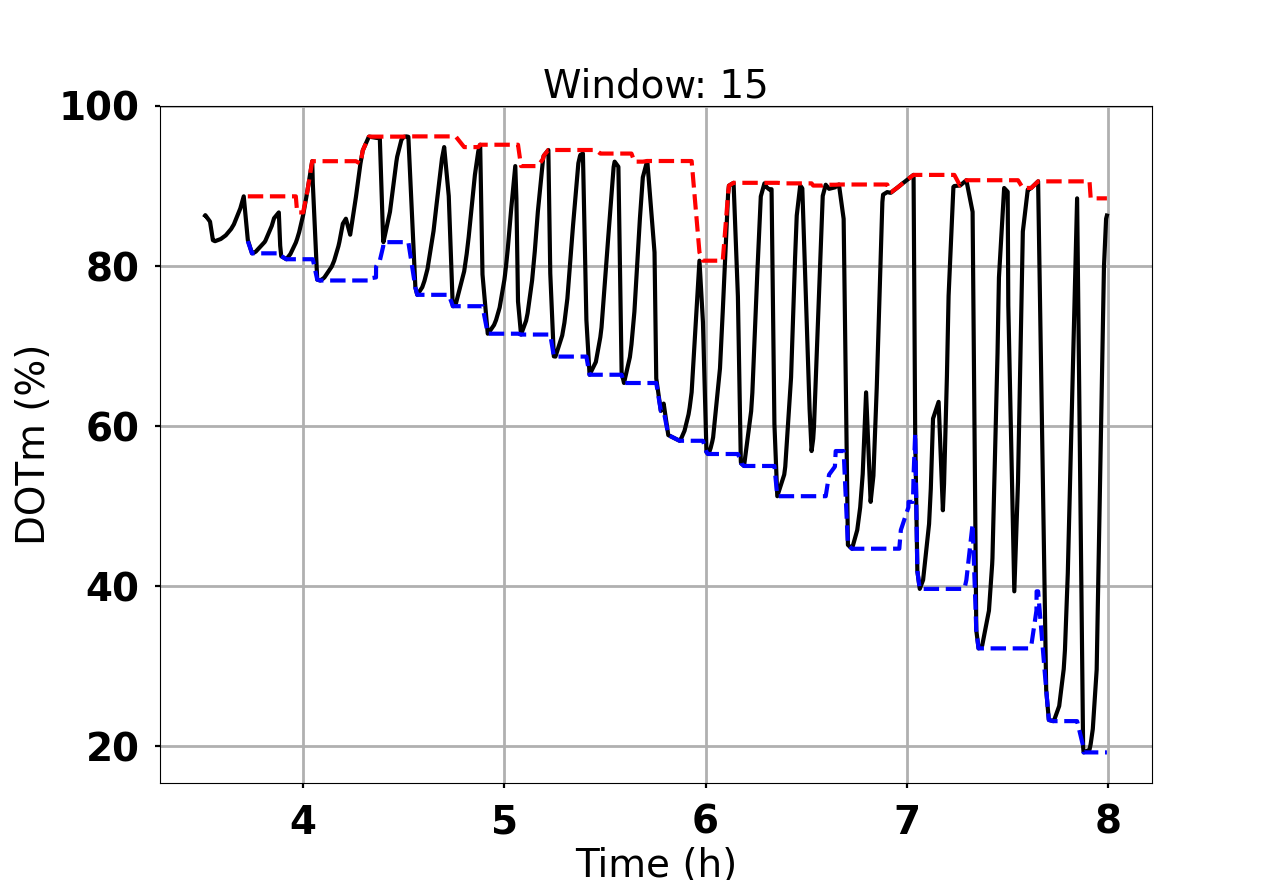} 
    \caption{Red: moving maximum of the signal in a 15 minute window; Blue: moving minimum with same window}
    \label{fig:Mmax_Mmin_w15}
\end{figure}

Where $w$ is the window size, $n_w$ is the total number of windows, $y^m$ is the measured state and $y$ is the simulated one. For the MAE of the moving minimum, the same definition holds. In figure \ref{fig:Mmax_Mmin_w15}, an example of the moving maxima and minima for a given signal is shown for a window of size 15, which having a sampling period of 60s implies windows of 15 minutes. If the smoothed envelope of the oxygen signal wants to be captured, the window size should be higher than the feeding period and smaller than two periods in order to properly capture the oxygen drops that are observed after the feed pulses.

In summary, the ACP will be higher when the MAE of the moving maxima and minima is smaller. This is intended to give a quantitative score of the goodness of the signal reconstruction, which is important for a correct monitoring and control of oxygen limitation in the culture as well as for having a better approximation of the oxygen uptake rate during the fast drop. On the other side, the estimated parameter values will also be compared with the \textit{true values} that generate the data. Even if the results are not that significant when not reporting a proper uncertainty quantification, this study aims to show the potential of this method for more detailed future analysis.

\section{Results}

\begin{figure}[h]
    \centering
    \includegraphics[trim={0 0 0 2cm}, clip, width=\linewidth]{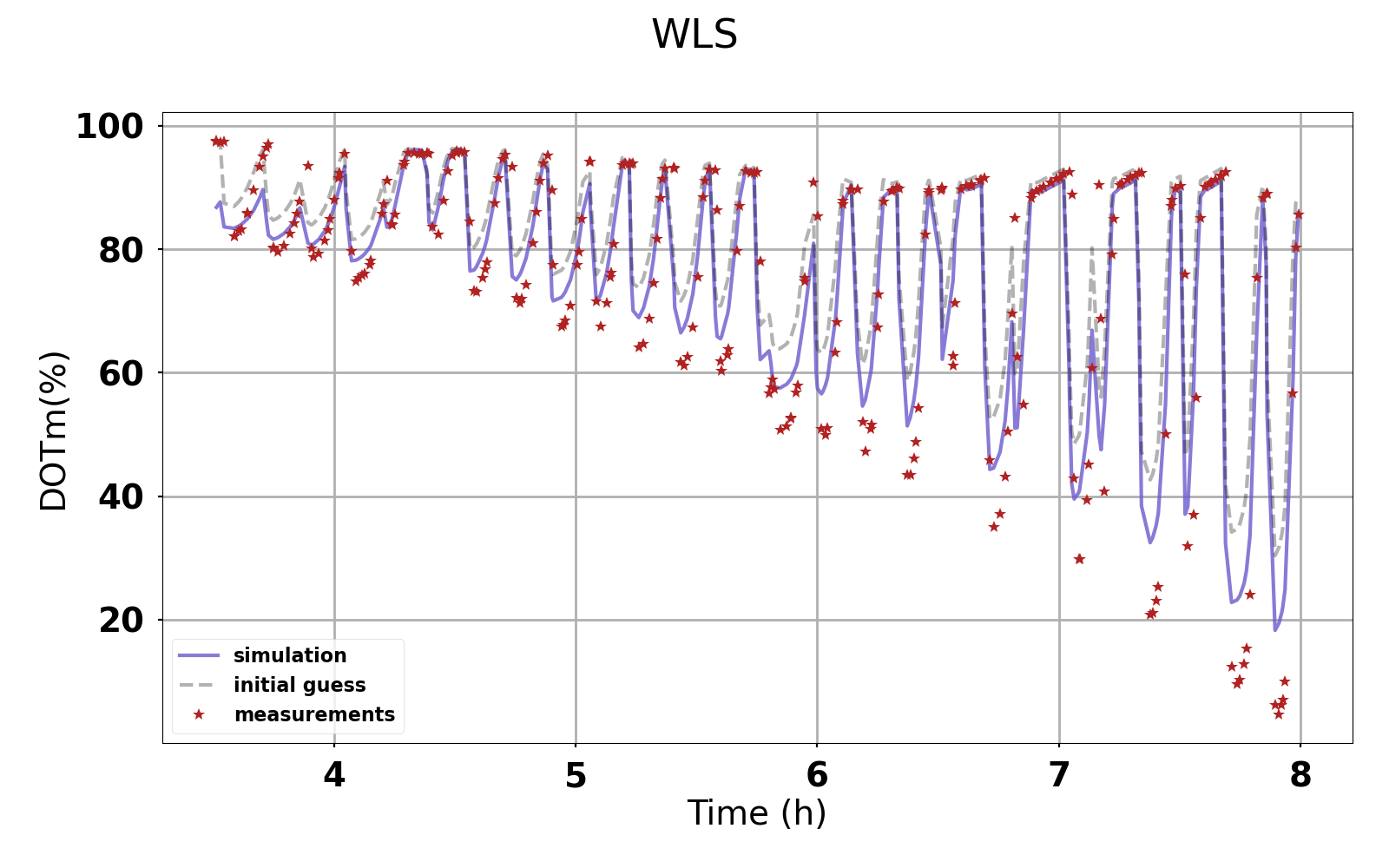}
    \caption{Results from fitting the model to the synthetic dataset using the WLS objective.}
    \label{fig:WLS_fit_WP}
\end{figure}

In figure \ref{fig:WLS_fit_WP}, the fed-batch phase fitting result for the WLS objective is shown, where a clear mismatch between synthetic data and simulation is observed at first glance, mostly for the low oxygen values corresponding to maximum oxygen uptake rate intervals. Together with the optimisation results, the simulation with the initial parameter guess is shown in dashed grey.

\begin{figure}[h]
    \centering
    \includegraphics[trim={0 0 0 2cm}, clip, width=\linewidth]{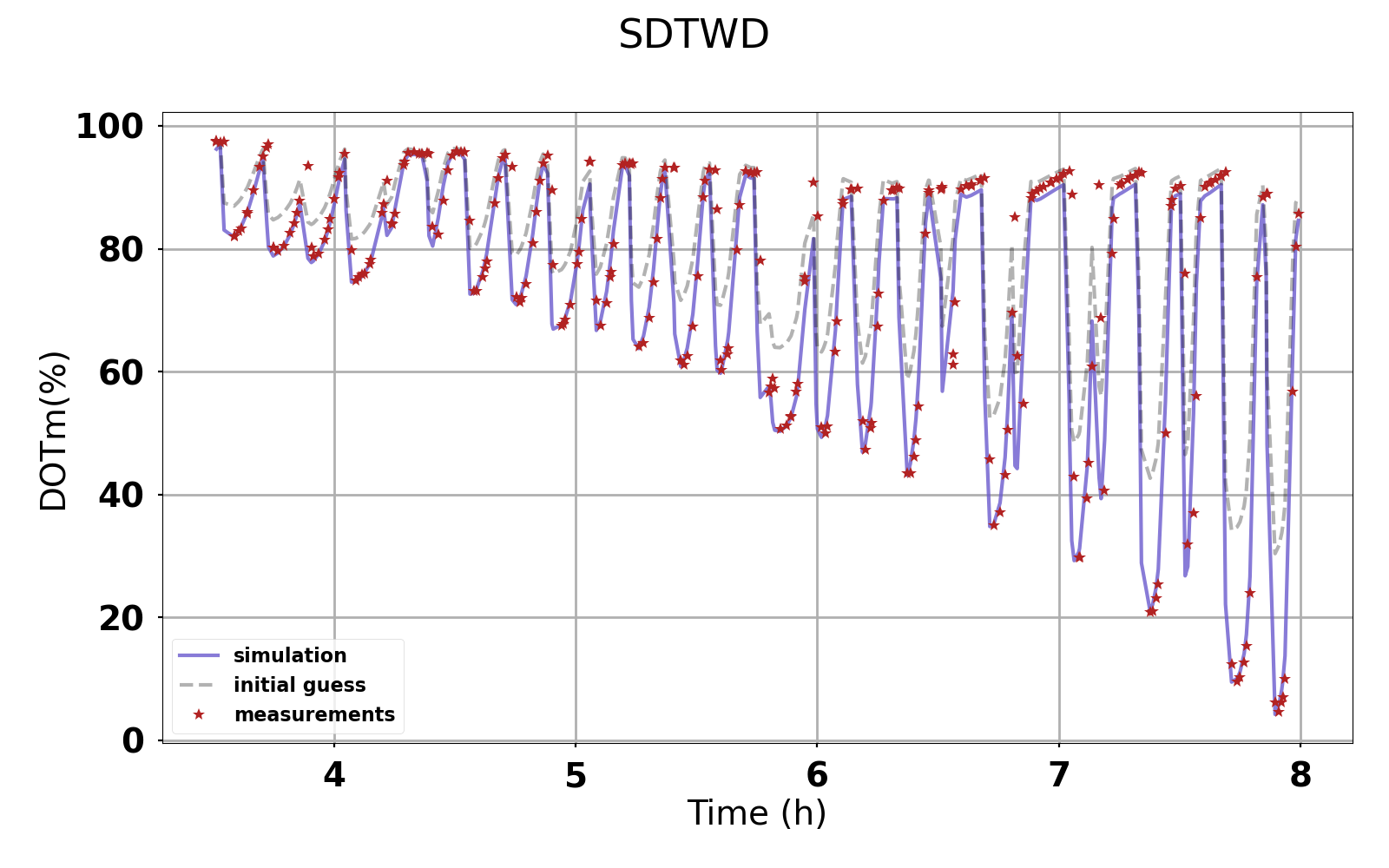}
    \caption{Results from fitting the model to the synthetic dataset using the SDTWD objective.}
    \label{fig:DTW_fit_WP}
\end{figure}

On the other side, figure \ref{fig:DTW_fit_WP} shows the results obtained using the SDTWD as objective function. In this case, due to the time invariance of this objective, the shape of the signal is well preserved and the oxygen drops can be properly fitted.

\begin{figure}[h]
    \centering
    \includegraphics[trim={0 0 2cm 0}, clip, width=\linewidth]{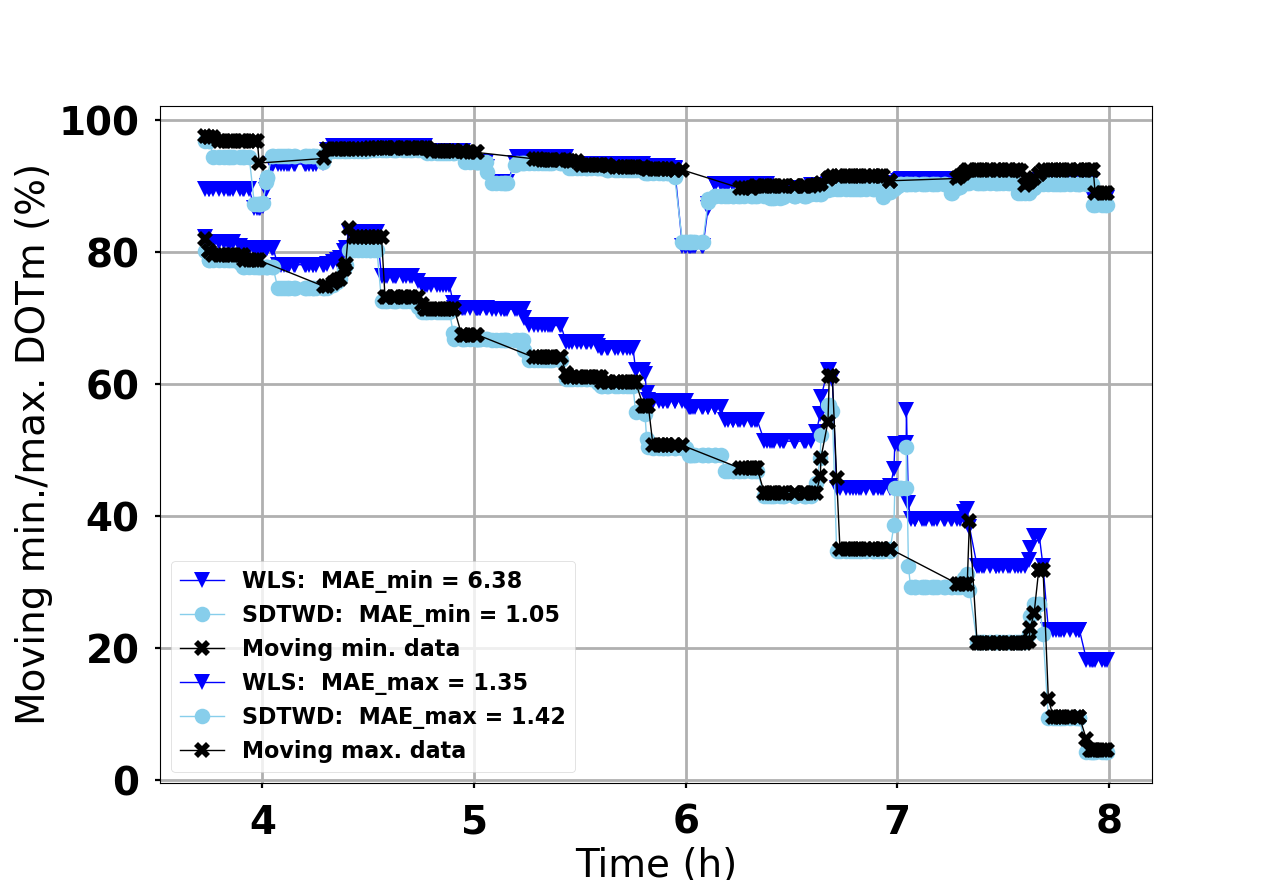}
    \caption{Moving maximums and minimums for the two estimation results and the data.}
    \label{fig:MAE_w15}
\end{figure}

The performance of the estimations when capturing the signals maximums and minimums is graphically represented in figure \ref{fig:MAE_w15}, where the dark blue markers show the amplitude of the signal estimated using WLS, the light blue corresponds to the SDTWD and the black to the synthetic data. The MAE for each of the results is shown in the legend, and as can be seen, the error when capturing minimums using the WLS estimation procedure is six times larger than when using SDTWD.

Finally, the estimated parameter values are show in table \ref{tb:estimated_parameters}, where the parameters estimated using the WLS objective are further from the truth than the ones given by the SDTWD. 

\begin{table}[h]
	\caption{True and estimated parameters with both objectives}
	\begin{center}
		\begin{tabular}{l c c c c c} \hline
            Parameter    &  Truth    &    WLS    &   SDTWD    \\ \hline
            $q_{smax}$   &   1.60    &    1.56   &    1.56    \\
            $Y_{xsem}$   &   0.59    &    0.50   &    0.53    \\
            $k_{L}a$     &   373.6   &    440.1  &    375.1   \\
            \hline 
		\end{tabular} 
	\end{center}
	\label{tb:estimated_parameters}
\end{table}

\section{Conclusion}
In this work, an insilico analysis of the effect that delayed inputs provoke in the oxygen signal reconstruction and model parameter estimation is presented. The experimental data has been generated containing the errors that are expected on the real system, where the delayed inputs are considered to imitate time shifts due to sensor delay, non-ideal mixing, and delayed inputs. The hypothesis that the Weighted Least Squares objective \textit{can fail} providing correct results in the presence of this time uncertainties is proved by an example with fast oxygen responses to the sudden substrate concentration changes. An alternative objective function that can deal with this kind of uncertainties is presented. The results using a differentiable divergence between time series based on the Dynamic Time Warping measure show superior performance on oxygen signal reconstruction and drop capturing, as well as less biased parameter estimates. In addition, as this analysis just considers fitting the oxygen signal for the model identification, the parameter estimates are expected to considerably improve when taking into account additional state measurements. 

One of the reasons why this proposed objective function performs properly in this case study, is the pattern-like dynamics that the oxygen manifests in response to the periodic bolus feeds. In cases where slow dynamics are present, this objective would not be appropriate, as sub-optimal minima can be found where faster or slower rate parameters generate the same output, and at the same time the standard estimation procedure could probably yield reasonable results. This problem might be solved by constraining the alignment path within a finite size window or the slope of the moves along the warping path, as is usually seen in applications of the classical Dynamic Time Warping.

Additionally, a new performance measure has been proposed in order to evaluate the goodness of the fit, the Amplitude Capturing Power. This measure offers a quantitative value for the fitting evaluation that is usually made by simple visual inspection, and in addition, is expected to yield better estimates for the oxygen uptake rates of the organisms. However, a closed form for this measure has not been defined yet.

As future work, a more general analysis which includes a better representation of the expected errors - as a time varying oxygen transfer rate - and a proper analysis of the uncertainty on the estimates is planned.


\bibliography{ifacconf}

\end{document}